\documentclass[12pt]{article}
\usepackage{amsmath,amsfonts,amssymb,epsfig}
\usepackage{graphicx}
\usepackage{pict2e}
\usepackage{times,setspace}

\newcommand{\staraco}[2]{\left\{ #1\stackrel{\star}{,}#2\right\} }
\newcommand{\sumint}[1]{\sum_{ #1} \!\!\!\!\!\!\!\int}
\def\be{\begin{equation}}
\def\ee{\end{equation}}
\def\bea{\begin{eqnarray}}
\def\eea{\end{eqnarray}}
\def\bb{\begin{equation*}}
\def\eb{\end{equation*}}
\def\beb{\begin{eqnarray*}}
\def\eeb{\end{eqnarray*}}
\def\re{{\rm e}}
\def\ri{{\rm i}}

\begin{document}

\thispagestyle{empty}
\begin{titlepage}

\begin{flushright}
UWThPh-2012-4
\\

\end{flushright}

\vspace{0.3cm}
\boldmath
\begin{flushleft}
  {\Large {\bf Degenerate noncommutativity }}
\unboldmath
\vspace{0.8cm}

{\bf
Harald~Grosse and Michael~Wohlgenannt
}

\end{flushleft}
\vskip 1em
\begin{flushleft}
University of Vienna, Faculty of Physics\\
Boltzmanngasse 5, A-1090 Vienna, Austria\\
Emails: harald.grosse, michael.wohlgenannt@univie.ac.at
\end{flushleft}

\vspace{6.6cm}
\begin{abstract}
\noindent We study a renormalizable four dimensional model with two deformed
quantized space directions. A one-loop renormalization is performed explicitly. The
Euclidean model is connected to the Minkowski version via an analytic continuation. At a
special value of the parameters a nontrivial fixed point of the renormalization group occurs.
\end{abstract}

\end{titlepage}


\section{Introduction}

During the last years a lot of efforts has been done to obtain renormalizable four dimensional
quantum field theory models defined over deformed space-time. These models suffer from the infrared-ultraviolet mixing. In a common work of one of us (H. G.) with Raimar Wulkenhaar \cite{Grosse:2003nw,Grosse:2004yu} we
realized, that a Lagrangian with four relevant/marginal operators leads to a renormalizable
model. In addition, a nontrivial fixed point occurs in parameter space, at which the
beta function for the coupling constant vanishes. Later on different renormalizable models were proposed,
\cite{VignesTourneret:2006nb, Gurau:2008vd, Grosse:2008df}. All these models are defined on a Euclidean deformed space.
There have been attempts to obtain similar results for models over deformed Minkowski
space-time, but the technical problems are, of course, much greater, and no final conclusion
has been obtained, up to now, see \cite{Fischer:2008dq}, \cite{Zahn:2010yt}, \cite{Bahns:2010dx} and \cite{Fischer:2011wi}.\\
In this note we consider canonical noncommutativity in two space-directions only,
\be
\label{m1}
\theta^{\mu\nu} = \left( \begin{array}{cccc} 0&0&0&0\\0&0&-\theta&0\\0&\theta&0&0\\0&0&0&0 \end{array}\right)\,.
\ee
A scalar field on such a deformed space has been shown to be renormalizable, if one takes
five operators into account. Since time is commutative, we expect that the model makes sense
also in deformed Minkowski space-time. \\
We also remark that we (one of us in a collaboration \cite{Grosse:2011es}) have been able to obtain conditions, which allow an analytic continuation from
Euclidean deformed space-time model to its Minkowski version in case of a commutative time coordinate
and under a condition on the time zero algebra.\\
We perform the one-loop renormalization explicitly and show that the renormalized amplitudes are
connected by an analytic continuation, and that in both cases a fixed point occurs.


\section{Euclidean space}

We are considering the model introduced in \cite{Grosse:2008df},
\bea
\label{euclidean:action-1}
S[\phi] & = & \int d^2xd^2y \frac 12 \phi(x,y)(-\Delta + \frac{4 \Omega^2}{\theta^2}y^2 + M^2) \phi(x,y)\\
\nonumber
&& + \frac{\kappa^2}{\theta^2} \int d^2xd^2yd^2z \phi(x,y)\phi(x,z)
+ \frac\lambda {4!} \int d^4x \phi(x)^{\star 4} \,,
\eea
where the second argument of $\phi$ refers to the noncommuting coordinates $y=(x_2,x_3)^T$. The term proportional to $\kappa^2$ is considered as an interaction, and $\kappa^2$ is a dimensionless coupling.
Using a matrix base for the noncommutative directions,
\begin{align}
\nonumber
f_{mn} = \frac 1{\sqrt{n!m!\theta^{m+n}}} \bar a^{\star m}  \star f_0 \star a^{\star n}\,,
\end{align}
where the ground state is given by $f_0 = 2 \re^{-\frac 1\theta (x_2^2+x_3^2)}$ and with $a=\frac 1{\sqrt{2}} (x_2 + \ri x_3)$ and $\bar a = \frac 1{\sqrt{2}} (x_2 - \ri x_3)$,
we can expand the fields in the following way:
\be
\phi(x,y) = \sum_{m,n} f_{mn} \phi_{mn}(x)\,,\quad
\phi_{mn}(x) = {\rm tr} f_{nm}\phi(x,y)\,,
\ee
and obtain for the action in momentum space
\begin{align}
\label{euclidean-action-2}
 S[\phi] = & 2\pi \theta \sum_{m,n,k,l} \int d^2Q \frac 12 \tilde \phi_{mn}(Q) \Delta^Q_{mn,kl}
\, \tilde \phi_{kl}(-Q)\\
\nonumber
& + (2\pi \theta)^2\,\frac{\kappa^2}{\theta^2} \sum_{m,n} \int d^2Q \tilde \phi_{mm}(Q) \tilde \phi_{nn}(-Q)\\
\nonumber
& + 2\pi\theta \frac\lambda {4!} \sum_{m,n,k,l} \int \prod_{i=1}^4 \frac {d^2 Q_i}{2\pi}
(2\pi)^2 \delta^{(2)}(Q_1+Q_2+Q_3+Q_4)\, \\
\nonumber
& \hspace{1.8cm}
\times \tilde \phi_{mn}(Q_1) \tilde \phi_{nk}(Q_2) \tilde \phi_{kl}(Q_3) \tilde \phi_{lm}(Q_4) \,.
\end{align}
The matrix elements for the kinetic part $\Delta^Q_{mn,kl}$ is diagonal in case of $\Omega=1$,
\begin{align}
\Delta^Q_{mn,kl} = (  Q^2 + M^2 + \frac 4\theta(m+n+1)) \delta_{nk}\delta_{ml}\,.
\end{align}
For $\Omega\ne 1$, also off-diagonal elements are present:
\begin{align}
\label{kernel-om-ne-1}
& \Delta^Q_{mn,kl} = (  Q^2 + M^2 + \frac {2(1+\Omega^2)}\theta(m+n+1) )\delta_{nk}\delta_{ml} \\
\nonumber
& - \frac{2(1-\Omega^2)}\theta \sqrt{(n+1)(m+1)} \delta_{n+1,k} \delta_{m+1,l}
 - \frac{2(1-\Omega^2)}\theta \sqrt{nm} \delta_{n-1,k} \delta_{m-1,l}
 \,.
\end{align}


\subsection{Feynman rules}

\paragraph{Propagator.}
\unitlength 1mm
We treat the noncommutative coordinates in the matrix base and the commutative ones in momentum base.
For a discussion of the matrix base see \cite{Grosse:2003nw,GraciaBondia:1987kw}.
\begin{itemize}
 \item For $\Omega=1$, we obtain
\bea
\nonumber
G_{m_1n_1,m_2n_2}(P,Q) & := & < \tilde \phi_{m_1n_1}(P_1,P_4) \tilde \phi_{m_2n_2}(Q_1,Q_4) > \\
& = &  \frac{\delta_{m_1n_2} \delta_{n_1m_2} \delta(P_1-Q_1) \delta(P_4-Q_4) }
{P_1^2 + P_4^2 + M^2 + \frac 4\theta(m_1+n_1 +1) }
\label{eucl-prop-Omega=1} \\
\nonumber
& \equiv & \delta^{(2)}(P-Q) G_{m_1n_1,m_2n_2}(P)\,.
\eea

\item For $\Omega\ne 1$, we can use the result of \cite{Rivasseau:2007ab}. We have to replace $\mu_0^2$ in (4.5) by
$$
\mu_0^2 \to P_1^2 + P_4^2 + M^2\,,
$$
using $D=2$, we obtain:
\begin{align}
  \label{eq:propaPhimatrix}
& G_{m, m+h; l + h, l} (P)
= \frac{\theta}{8\Omega} \int_0^1 d\alpha\,
\frac{(1-\alpha)^{\frac{ ( P_1^2 + P_4^2 + M^2 ) \theta}{8 \Omega}-\frac{1}{2}}}{1 + C\alpha } G^{(\alpha)}_{m, m+h; l + h, l},
\\
\label{G**alpha}
& \hspace{-1cm}
G^{(\alpha)}_{m, m+h; l + h, l}
= \left(\frac{\sqrt{1-\alpha}}{1+C \alpha}
\right)^{m+l+h} \sum_{u=\max(0,-h)}^{\min(m,l)}
   {\mathcal A}(m,l,h,u)
\left( \frac{C \alpha (1+\Omega)}{\sqrt{1-\alpha}(1-\Omega)}
\right)^{m+l-2u},
\end{align}
where ${\mathcal A}(m,l,h,u)=\sqrt{\binom{m}{m-u}
\binom{m+h}{m-u}\binom{l}{l-u}\binom{l+h}{l-u}}$, and $C$ is a function of $\Omega$,
$$
C(\Omega) = \frac{(1-\Omega)^2}{4\Omega}\,.
$$

\end{itemize}

\paragraph{Vertex weights.}

\begin{align}
  & \parbox{20mm}{\begin{picture}(20,15)
       \put(0,0){\epsfig{file=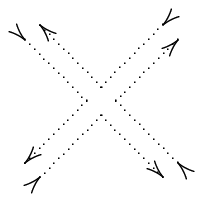,bb=71 638 117 684}}
   \put(-4,12){\mbox{\footnotesize$n_4$}}
       \put(-1,6){\mbox{\footnotesize$m_1$}}
       \put(13,9){\mbox{\footnotesize$m_3$}}
       \put(14,4){\mbox{\footnotesize$n_2$}}
       \put(4,-1){\mbox{\footnotesize$n_1$}}
       \put(10,-1){\mbox{\footnotesize$m_2$}}
       \put(5,14){\mbox{\footnotesize$m_4$}}
       \put(10,15){\mbox{\footnotesize$n_3$}}
   \end{picture}}
\qquad = - \lambda \delta^{(2)}(P_1+P_2+P_3+P_4) \delta_{n_1m_2}\delta_{n_2m_3}\delta_{n_3m_4}\delta_{n_4m_1}
\label{vertex}
\end{align}

\begin{align}
  & \parbox{20mm}{\begin{picture}(10,15)
       \put(0,7){\mbox{\footnotesize$\mathbf P_1$}}\put(4,7){\vector(1,0){7}}
       \put(26,7){\vector(1,0){7}}\put(34,7){\mbox{\footnotesize$\mathbf P_2$}}
       \put(12,5){\line(1,0){5}}\put(12,9){\line(1,0){5}}
       \put(17,6.9){\oval(4,4)[r]}\put(24,6.9){\oval(4,4)[l]}
       \put(24,5){\line(1,0){5}}\put(24,9){\line(1,0){5}}
       \put(8,3){\mbox{\footnotesize $n$}}\put(8,8){\mbox{\footnotesize $m$}}
       \put(31,3){\mbox{\footnotesize $k$}}\put(31,8){\mbox{\footnotesize $l$}}
   \end{picture}}
\qquad \qquad \, = - 2 \, \frac{\kappa^2}{\theta^2} \delta_{mn} \delta_{kl} \delta^{(2)}(P_1 - P_2)
\label{vertex-counter}
\end{align}

\paragraph{Power counting.}
In \cite{Grosse:2008df}, the power counting degree of convergence for a Feynman graph $G$ on a Riemann surface with genus $g$, boundary components $b$ and for $\kappa=0$ was derived:
\be
\label{power-counting}
\omega(G) = N-4+4g + 2(b-1)\,.
\ee
For comparison, in the non-degenerate case the degree of convergence is given by
\begin{align}
\nonumber
\omega_{nd} (G) = N-4+8g + 4(b-1)\,.
\end{align}
This explains the necessity of the fifth operator in the defining action, since a graph with $N=2$, $g=0$ and $b=2$
is logarithmic divergent in the former (see \eqref{nonplanar}) but finite in the latter case.


\subsection{1-loop calculation}

In this section, we are computing all the divergent 1-loop diagrams according to the power counting formula \eqref{power-counting}.

\subsubsection{$\Omega=1$}

Let us first consider the case $\Omega=1$. The first divergent contribution is the planar, regular two-point tadpole.

\begin{align}
  & \parbox{25mm}{\begin{picture}(22,30)
      \put(0,0){\epsfig{file=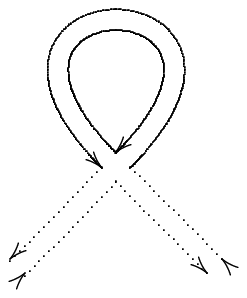}}
      \put(-1,6){\mbox{\scriptsize $m_1$}}
      \put(3,0){\mbox{\scriptsize $n_1$}}
      \put(14,2){\mbox{\scriptsize $m_2$}}
      \put(21,4){\mbox{\scriptsize $n_2$}}
      \put(10,16){\mbox{\scriptsize $l$}}
\end{picture}}
+~~ \parbox{25mm}{\begin{picture}(22,30)
    \put(0,0){\epsfig{file=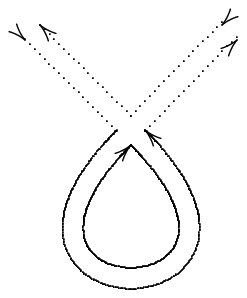}}
    \put(-1,22){\mbox{\scriptsize $n_1$}}
    \put(4,26){\mbox{\scriptsize $m_1$}}
    \put(16,28){\mbox{\scriptsize $n_2$}}
    \put(21,22){\mbox{\scriptsize $m_2$}}
    \put(10,11){\mbox{\scriptsize$l$}}
\end{picture}}
\qquad =  T_1 = T_{\rm up} + T_{\rm down} \nonumber
\\
\nonumber
& =
- \delta_{m_1n_2} \delta_{m_2n_1} \sumint{\,\,\, l}d^2P
\bigg( \frac \lambda {P^2 + \frac 4\theta(l+m_1+1)+M^2} \\
& \hspace{1cm}
+ \frac \lambda {P^2 + \frac 4\theta(l+m_2+1)+M^2} \bigg)
\nonumber
\\
\label{r1a}
& = - \delta_{m_1n_2} \delta_{m_2n_1} \, \frac{\pi\theta\lambda}4 \, \Bigg(
\frac 2{\epsilon} +  ( \frac 4\theta ( m_1 + m_2 + 1) + 2M^2 ) \ln \epsilon \Bigg)
+ \mathcal O(\epsilon^0)\,.
\end{align}
The symmetry factor for $T_1$ is $\frac 16$. The above logarithmic divergence renormalizes the indices $m$ and $n$, and
also the mass $M^2$. This is a wave function
renormalization at 1-loop. But the ``commutative'' part of the kinetic term, $Q^2$, is not renormalized (only at two loops, sunset diagram).

As we have seen above, the planar, non-regular two-point tadpole with two boundary components is logarithmically divergent.
\begin{align}
&\parbox{20\unitlength}{\begin{picture}(50,50)
       \put(5,23){\line(1,0){10}}\put(18,23){\line(1,0){10}}
       \put(5,26){\line(1,0){10}}\put(18,26){\line(1,0){10}}
       \put(0,20){$m$}\put(0,25){$l$}
       \put(26,23){\oval(22,22)[b]}\put(26,23){\oval(16,16)[b]}
       \put(26,26){\oval(22,22)[t]}\put(26,26){\oval(16,16)[t]}
       \put(37,23){\line(0,1){3}}\put(34,23){\line(0,1){3}}
       \put(29,20){$n$}\put(29,26){$k$}
       \put(21,39){\vector(1,0){8}}\put(23,42){$\mathbf p$}
\end{picture}}
\nonumber
\quad\qquad\qquad = T_2\\
\nonumber
&
= \int_\Lambda d^2p \frac{-\lambda \delta_{lm}\delta_{kn}}{p^2+\frac 4\theta (l+k+1)+M^2}
\\
\label{nonplanar}
& = \delta_{lm}\delta_{kn}\, \lambda \pi\left( \ln \epsilon + \gamma + \ln(\frac 4\theta (l+k+1)+M^2) \right) + \mathcal O(\epsilon^0)\,.
\end{align}
The symmetry factor for $T_2$ is $\frac 16$.
The divergent part is independent of the oscillator indices. Therefore, the singularity can easily be
subtracted by choosing any fixed indices $l_0$ and $k_0$. The values should be fixed by normalization conditions.

The last diagram we have to consider is the following 4-point graph:

\begin{align}
& \parbox{50mm}{\begin{picture}(50,15)
       \put(12,0){\epsfig{file=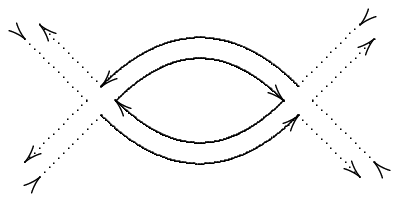,bb=71 630 174 676}}
       \put(8,12){\mbox{\footnotesize$m$}}
       \put(11,6){\mbox{\footnotesize$m$}}
       \put(46,9){\mbox{\footnotesize$k$}}
       \put(48,4){\mbox{\footnotesize$k$}}
       \put(16,-1){\mbox{\footnotesize$n$}}
       \put(42,-1){\mbox{\footnotesize$n$}}
       \put(17,14){\mbox{\footnotesize$l$}}
       \put(42,15){\mbox{\footnotesize$l$}}
       \put(24,8){\mbox{\footnotesize$r$}}
       \put(34,8){\mbox{\footnotesize$r$}}
       \put(0,8){\mbox{\footnotesize $P$}}\put(3,8){\vector(1,0){5}}
       \put(45,8){\vector(1,0){5}}\put(52,8){\mbox{\footnotesize $P$}}
       \put(29,17){\mbox{\footnotesize$Q$}}\put(26,-2){\mbox{\footnotesize$P+Q$}}
       \put(32,16){\vector(-1,0){4}}\put(27,1){\vector(1,0){6}}
   \end{picture}}
\qquad = T_3
\nonumber\\
\nonumber
& = \lambda^2\,\sumint{\,\,\,r}_{\,\,\,\Lambda} d^2 Q \frac 1{Q^2+\frac 4\theta (l+r+1)+M^2}\,
\frac 1{(P+Q)^2+\frac 4\theta (n+r+1)+M^2} \\
\label{A4diff}
& = - \lambda^2 \frac{\theta\pi}{4} \ln\epsilon + \mathcal O(\epsilon^0)\,.
\end{align}
The symmetry factor for $T_3$ is $\frac 1{3}$.

\subsubsection{\label{Om-2-point} $\Omega\ne 1$}

In the general case $\Omega\ne1$, the computation of the contribution of the same diagrams is more involved.
The planar, regular tadpole is given by

\begin{align}
  & \parbox{25mm}{\begin{picture}(22,30)
      \put(0,0){\epsfig{file=a12.eps}}
      \put(-1,6){\mbox{\scriptsize $m_1$}}
      \put(3,0){\mbox{\scriptsize $n_1$}}
      \put(14,2){\mbox{\scriptsize $m_2$}}
      \put(21,4){\mbox{\scriptsize $n_2$}}
      \put(10,16){\mbox{\scriptsize $l$}}
\end{picture}}
+~~ \parbox{25mm}{\begin{picture}(22,30)
    \put(0,0){\epsfig{file=a12a.eps}}
    \put(-1,22){\mbox{\scriptsize $n_1$}}
    \put(4,26){\mbox{\scriptsize $m_1$}}
    \put(16,28){\mbox{\scriptsize $n_2$}}
    \put(21,22){\mbox{\scriptsize $m_2$}}
    \put(10,11){\mbox{\scriptsize$l$}}
\end{picture}}
\qquad =  T^\Omega_1 = T^\Omega_{\rm up} + T^\Omega_{\rm down}\,,
\nonumber
\end{align}
where
\begin{align}
T_{\rm up}^\Omega & = -
\lambda \delta^{(2)}(P_1-P_2) \delta_{m_1n_2} \delta_{m_2n_1} \sum_{k} \int d^2Q \, G(Q)_{m_1k,k m_1}
\end{align}
and
\begin{align}
T_{\rm down}^\Omega & = -
\lambda \delta^{(2)}(P_1-P_2) \delta_{m_1n_2} \delta_{m_2n_1} \sum_{k} \int d^2Q \, G(Q)_{k n_1,n_1 k}
\,.
\end{align}
Let us consider $T^\Omega_{\rm up}$ first. For the propagator, we use Eq. \eqref{eq:propaPhimatrix}:
\begin{align}
G(Q)_{m_1l,lm_1} = & \frac \theta{8\Omega} \int_0^1 d\alpha  \frac{ (1-\alpha)^{(Q^2+M^2)\frac\theta{8\Omega} - \frac12} }
{ 1 + C \alpha } \left(  \frac{1-\alpha}{(1+C\alpha)^2}  \right)^{\frac 12 (m_1 + l)}\\
\nonumber
& \sum_{u={\rm max}(0,m_1 - l)}^{m_1} \binom{m_1}{m_1 - u} \binom{l}{m_1 - u}
\left( \frac{ C \alpha(1+\Omega) }{ \sqrt{1-\alpha} (1-\Omega) }  \right)^{2(m_1 - u)}
\end{align}
and the substitution $(1-\alpha) = e^{-\beta \frac{8\Omega}\theta}$.

We have to analyse the sum over $u$ in more detail. We only concentrate on the divergent contributions. For simplicity, we use $m$ instead of $m_1$.
\begin{itemize}
\item \underline{$u=m$:}
\begin{align}
T_{\rm up}^{\Omega,m} & = \pi \sum_{k=0}^\infty \int_\epsilon^\infty \frac {d\beta}\beta
\frac{ \re^{-\beta \frac{4\Omega k}\theta} } { ( 1+C(1-\re^{-\beta\frac{8\Omega}\theta}) )^{m+k+1} }
\re^{-\beta(M^2 + \frac{4\Omega}\theta (1+m))}\\
\nonumber
& = \pi \int_\epsilon^\infty \frac {d\beta}\beta
\frac{ ( 1+C(1-\re^{-\beta\frac{8\Omega}\theta}) )^{-m} \re^{\beta \frac{4\Omega}\theta} }
{ \re^{\beta \frac{4\Omega}\theta} - 1 + C(1-\re^{-\beta\frac{8\Omega}\theta}) \re^{\beta \frac{4\Omega}\theta}}
\re^{-\beta(M^2 + \frac{4\Omega}\theta (1+m))}\,.
\end{align}
In the next step, we expand the integrand -- except for the exponential in order to avoid ordinary IR divergences -- for small values of $\beta$, since we are only interested in the divergent contributions and small values of $\beta$ correspond to large momenta.
\begin{align}
T_{\rm up}^{\Omega,m} & =
 \frac \pi{1+2C} \left(
\frac\theta{4\Omega} \left( \frac 1\epsilon + \ln\epsilon (\frac{4\Omega}\theta(m+1) + M^2) \right)
\right.\\
& \hspace{3.4cm} \left.
- \frac{ 1 + 4C }{2(1+2C)} \ln\epsilon + 2Cm \ln\epsilon
\right) + \mathcal O(\epsilon^0)
\end{align}

\item \underline{$u=m-1$:}
\begin{align}
T_{\rm up}^{\Omega,m-1} & =\pi \sum_{k=1}^\infty \int_\epsilon^\infty \frac {d\beta}\beta
\binom{m}{1} \binom{k}{1}\frac{ (1-\re^{-\beta\frac{8\Omega}\theta})^{2}\re^{-\beta \frac{4\Omega k}\theta} }
{ ( 1+C(1-\re^{-\beta\frac{8\Omega}\theta}) )^{m+k+1} } \\
\nonumber
&\times
\left( \frac{(1+\Omega)(1-\Omega)}{4\Omega} \right)^{2} \re^{-\beta(M^2 + \frac{4\Omega}\theta (m-1))}\\
\nonumber
& = - \frac{ 4 \pi m }{(1+2C)^2} \left( \frac{(1+\Omega)(1-\Omega)}{4\Omega} \right)^{2} \ln\epsilon +\mathcal O(\epsilon^0)\,.
\end{align}

\item \underline{$u=m-2$:}
This yields already a finite contribution. The same is true for smaller $u$'s, since the numerator of the integrand is proportional to $\beta^{2(m-u)}$, whereas the denominator, after summing over $k$, is proportional to $\beta^{m-u+2}$.

\end{itemize}

For $T^\Omega_{\rm down}$, we obtain the same result. This amounts to
\begin{align}
T^\Omega_1 & = -
\lambda \delta^{(2)}(P_1-P_2) \delta_{m_1n_2} \delta_{m_2n_1}
\Bigg\{
\frac \pi{1+2C} \left( \frac\theta{2\Omega \epsilon} + \frac{M^2 \theta}{2\Omega} \ln\epsilon \right)
\\
& \hspace{3cm}
+ \frac \pi{(1+2C)^2} ( m_1 + n_1 + 1 ) \ln \epsilon \Bigg\}
\end{align}
and is consistent with the one previously obtained in \eqref{r1a}.

\begin{align}
&\parbox{20\unitlength}{\begin{picture}(50,50)
       \put(5,23){\line(1,0){10}}\put(18,23){\line(1,0){10}}
       \put(5,26){\line(1,0){10}}\put(18,26){\line(1,0){10}}
       \put(0,20){$m$}\put(0,25){$l$}
       \put(26,23){\oval(22,22)[b]}\put(26,23){\oval(16,16)[b]}
       \put(26,26){\oval(22,22)[t]}\put(26,26){\oval(16,16)[t]}
       \put(37,23){\line(0,1){3}}\put(34,23){\line(0,1){3}}
       \put(29,20){$n$}\put(29,26){$k$}
       \put(21,39){\vector(1,0){8}}\put(23,42){$\mathbf p$}
\end{picture}}
\nonumber
\quad\qquad\qquad = T_2^\Omega = - \lambda \, \delta^{(2)} (P_1- P_2) \, \int d^2Q\, G_{kl;mn}(Q)\,,
\end{align}
where $l=k+h$ and $m=n+h$, for some $h$. We then obtain
\begin{align}
\nonumber
T_2^\Omega & = -\lambda \delta^{(2)}(P_1-P_2)
\frac{\theta}{8\Omega} \int_0^1 d\alpha\, \int d^2Q\,
\dfrac{(1-\alpha)^{\frac{ ( Q^2 + M^2 ) \theta}{8 \Omega}-\frac{1}{2}}}{1 + C\alpha }
\left(\frac{\sqrt{1-\alpha}}{1+C \alpha} \right)^{k+n+h} \\
\nonumber
& \times
\sum_{u=\max(0,-h)}^{\min(k,n)}
   {\mathcal A}(k,n,h,u)
\left( \frac{C \alpha (1+\Omega)}{\sqrt{1-\alpha}(1-\Omega)}
\right)^{k+n-2u}\\
\nonumber
& = \lambda \pi \delta_{kn} \delta_{lm} \delta^{(2)}(P_1-P_2) \ln\epsilon
\end{align}

The contribution of the 4-point graph depicted in \eqref{A4diff} yields
\begin{align}
T_3^\Omega & = - \frac{\lambda^2 \pi\theta}{2(1+\Omega^2)} \ln\epsilon\,.
\end{align}


\subsection{Explicit renormalization}

In this section, we explicitly perform the 1-loop renormalization using the results computed in the previous
subsections. Let us consider the general case, $\Omega\ne 1$.
In general, the 1-loop corrections are diagonal (cf. \cite{Grosse:2004by}) and the effective action
is renormalized according to
$$
S[{\mathcal Z}^{1/2} \tilde \phi]_{\rm 1-loop} =  S[\tilde \phi; M_{\rm phys},\lambda_{\rm phys}, \Omega_{\rm phys}, \kappa_{\rm phys}, a_{\rm phys}]\,,
$$
we obtain identifications for the physical parameters. In the commutative case, we have no wave function renormalization at 1-loop.
The sunset diagram is at second order and the first diagram to introduce such a renormalization. In the noncommuative case, there
is a different picture. For non-degenerate noncommutativity, the planar two-point tadpole,
introduces a wave function
renormalization already at 1-loop order. In the case we are considering here, where the noncommutativity is restricted to two (spatial)
dimensions, we get a mixture of both scenarios. The 2-point tadpole renormalizes the wave function partly and breaks the symmetry between
commutative and noncommutative directions. In order to capture this asymmetry
we introduce an additional parameter $a$ for the
commutative directions. An alternative solution would be to scale the noncommutativity parameter $\theta$ instead.

The physical (renormalized) paramters can now be identified. The logarithmic derivative of the
inverse map with respect to the logarithm of the momentum cut-off $\Lambda = 1/ \sqrt{\epsilon}$
determines  $\beta$ functions:
\begin{align}
\nonumber
\beta_\Omega & = \frac{\lambda_{\rm phys} \Omega_{\rm phys} (1-\Omega_{\rm phys}^2)}{6(1+\Omega_{\rm phys}^2)^2} \,,\\
\beta_\lambda & =  \lambda_{\rm phys}^2 \frac{1-\Omega_{\rm phys}^2}{6(1+\Omega_{\rm phys}^2)^2} \,,\\
\nonumber
\beta_a & = - \frac{a_{\rm phys} \lambda_{\rm phys}\, \Omega_{\rm phys}^2}{6(1+\Omega_{\rm phys}^2)^2} \,,\\
\nonumber
\beta_{\kappa^2} & = - \frac{\kappa^2_{\rm phys} \lambda_{\rm phys} \Omega_{\rm phys}^2}{6(1+\Omega_{\rm phys}^2)^2}
- \frac{\pi \lambda_{\rm phys}}{3(2\pi)^2} \,,\\
\nonumber
\beta_{M^2} & =  - \lambda_{\rm phys} ( \frac{\Omega_{\rm phys}^2}{6(1+\Omega_{\rm phys}^2)^2} -
\frac 1{3(1+\Omega_{\rm phys})^2} ) \\
\nonumber
& \hspace{2cm}
 - \frac {\lambda_{\rm phys}}{6 M_{\rm phys}^2(1+\Omega_{\rm phys})^2} \Lambda^2 \,.
\end{align}

\subsubsection{$\Omega=1$}

The results for $\Omega=1$ coincide with those above, in the limit $\Omega\to 1$. It is remarkable that the physical coupling $\lambda_{\rm phys}$ is unchanged, i.e. it has a vanishing $\beta$ function, as in the non-degenerate case. Probably, this effect does not sustain to higher order loop corrections, because of the asymmetry between commutative and noncommutative directions and the need to introduce the additional parameter $a$.
In the renormalisation proof (to all orders) in \cite{Grosse:2008df} using multiscale analysis, it was not necessary to introduce this parameter $a$. Somehow, multiscale analysis seems to be unaffected or insensitive of this asymmetry, which appears in the explicit calculations here!


\section{Minkowski space-time}

For the deformed Minkowski space-time we shall restrict ourselves to $\Omega=1$. The kernel
for the quadratic part becomes

\begin{align}
\Delta^{M,Q}_{mn,kl} = (-Q_0^2 + Q_1^2 + M^2 + \frac 4\theta(m+n+1)) \delta_{nk}\delta_{ml}
\end{align}

We expand the scalar field in terms of plane waves - with respect to variables $t$ and $z$ - and the oscillator basis - with respect to $x$ and $y$:
\begin{align}
\label{m2}
\Phi(t,z) & = \sumint{mn} \frac{dp}{(2\pi)2\omega_{p,mn}} \bigg(
\re^{-\ri\, \omega_{p,mn} t} \re^{  \ri pz} f_{mn} \otimes a_{p,mn} \\
\nonumber
& \hspace{1.9cm} + \re^{ \ri\, \omega_{p,mn} t} \re^{- \ri pz} f_{nm} \otimes a^\dagger_{p,mn}
\bigg)\,,
\end{align}
with
\be
\label{m3}
\omega_{p,mn}^2 = p^2 + \frac 4\theta (m+n+1) + M^2\,.\,\, m,n\in \mathbb N\,,
\ee
\be
\label{m4}
\phi_{mn}(t,z) = \int \frac{dp}{(2\pi)2\omega_{p,mn}} \left(
\re^{-\ri\, \omega_{p,mn} t} \re^{  \ri pz} a_{p,mn} +
\re^{ \ri\, \omega_{p,nm} t} \re^{- \ri pz} a^\dagger_{p,nm}
\right)\,,
\ee
where
\be
\Phi(t,z) = \sum_{mn} f_{mn} \phi_{mn}(t,z)\,,\quad
\phi_{mn}(t,z) = {\rm tr} f_{nm}\Phi(t,z)\,.
\ee
This is a solution of the noncommutative wave equation:
\be
\partial_0^2 \Phi - \partial_z^2 \Phi - \staraco{\tilde y^2}{\Phi} + M^2 \Phi = 0\,,
\ee
where $y=(x_1,x_2)^T$ and $z=x_3$.
The creation and annihilation operators satisfy the following relations:
\begin{flushleft}
\bea
\nonumber
&& [a_{p,mn}, a^\dagger_{q,m'n'}] = (2\pi)\, 2 \omega_{p,mn}\,\delta(p-q) \delta_{mm'} \delta_{nn'}\,,\\
\nonumber
&& [a_{p,mn}, a_{q,m'n'}] = [a^\dagger_{p,mn}, a^\dagger_{q,m'n'}] = 0\,.
\label{m5}
\eea
\end{flushleft}
These relations lead to
\be
\label{m6}
[\phi_{mn}(t,z_1), \dot \phi_{m'n'}(t,z_2)] = \ri \delta(z_1 - z_2) \delta_{mn'}\delta_{nm'}\,,
\ee
and
\bea
\nonumber
[\phi_{mn}(t_1,z_1),\, \phi_{m'n'}(t_2,z_2)] & = & \\
\label{causality}
&& \hspace{-6cm}
= \int \frac{dp}{(2\pi)\,2\omega_{p,mn}}
\Bigg( \re^{-\ri\omega_{p,mn}(t_1-t_2)} \re^{\ri p(z_1-z_2)} - \re^{\ri\omega_{p,mn}(t_1-t_2)} \re^{-\ri p(z_1-z_2)}
\Bigg) \delta_{mn'}\delta_{nm'}\,.
\eea
The above Equation implies causality in the two-dimensional commutative subspace.

The Hamiltonian is given by:
\bea
H & = & \frac 12 \int d^3x \left(
(\partial_0 \Phi)^2 + (\partial_z \Phi)^2 + \Phi \staraco{\tilde x_1^2 + \tilde x_2^2}{\Phi} +M^2\, \Phi^2
\right) \\
\nonumber
& \equiv & \frac 12 \int d^3\tilde p \omega_{p,mn} \left( a^\dagger_{p,mn} a_{p,mn} + a_{p,mn} a^\dagger_{p,mn} \right)\,,
\eea
where we defined the measure $\int d^3\tilde p = 2\pi \theta \sum_{m,n} \int \frac{dp}{2\pi (2\omega_{p,mn})}$. Naturally, we can
define a normal ordering:
\be
:H: = \int d^3\tilde p\, \omega_{p,mn} a^\dagger_{p,mn} a_{p,mn} \,.
\ee
We have to study the time ordered vacuum expectation value of the product of $n$ fields:
\bea
\nonumber
(0|T \Phi \dots \Phi|0) & = & (0|T \sum_{k,l} f_{k_1l_n} \phi_{k_1k_2}\phi_{k_2k_3}\dots \phi_{k_nl_n}|0) \\
& = & \sum_{k,l} <f_{k_1l_n}><0|T\phi_{k_1k_2}\phi_{k_2k_3}\dots \phi_{k_nl_n}|0>\\
\nonumber
& = & \sum_{k,l} <0|T\phi_{k_1k_2}\phi_{k_2k_3}\dots \phi_{k_nk_1}|0> \,.
\eea
Using \eqref{m4}, we obtain
\bea
& <0| T \phi_{m_1n_1}(t_1,z_1) \phi_{m_2n_2}(t_2,z_2) |0> =
\int \frac{dp}{(2\pi)^2 2 \omega_{p,m_1n_1}} \delta_{m_1m_2}\delta_{n_1n_2} \\
& \nonumber
\times \Big(
\theta(t_1-t_2) \re^{\ri \omega_{p,m_1n_1}(t_1-t_2)-\ri p(z_1-z_2)} +
\theta(t_2-t_1) \re^{-\ri \omega_{p,m_1n_1}(t_1-t_2)+\ri p(z_1-z_2)}
\Big)\,.
\eea
More general, the expressions to be studied are expectation values of the form
\be
<0|\phi_{m_1n_1}(t_1,z_1)\phi_{m_2n_2}(t_2,z_2)\dots \phi_{m_Nn_N}(t_N,z_N)|0>\,,
\ee
which are in the undeformed case connected to the Euclidean correlation
functions by analytic continuation, see \ref{analytic-cont}.
Serious difficulties arise, if one considers a noncommutativity of the time coordinate as discussed in \cite{Bahns:2009iq}.

\subsection{Feynman rules}

The action for the scalar field on the deformed Minkowski space-time is given by
the Minkowski version of equation \eqref{euclidean-action-2}.
Therefore, the propagator reads
\begin{align}
\label{minkowski-prop}
G^M_{m_1n_1;m_2n_2}(P) = \frac{\ri}{-P_0^2+P_1^2 + \frac 4\theta (m_1+n_1 + 1) + M^2 + \ri \epsilon}\,,
\end{align}
whereas the vertex weight is given by
\begin{align}
\label{minkowski-vertex}
-\ri \lambda\, \delta_{n_1m_2}\delta_{n_2m_3}\delta_{n_3m_4}\delta_{n_4m_1} \delta^{(2)}(P_1+P_2+P_3+P_4)
\,.
\end{align}


\subsection{1-loop calculations}

The calculation of the divergent contributions of the above mentioned diagrams uses the Feynman rules \eqref{minkowski-prop} and \eqref{minkowski-vertex} and follows the same lines as in the Euclidean case.

\subsubsection{\label{renorm-mink}Explicit renormalization}

Similar as before, we obtain an effective action and the one-loop corrections in Minkowski space. The divergent corrections of the parameters in the Minkowski case coincide with the corresponding Euclidean ones. Therefore, the beta functions coincide as well. The most important result is the vanishing of the beta function of the coupling $\lambda$ at $\Omega = 1$.


\subsection{Analytic continuation}
\label{analytic-cont}

Let us start with the Wightman two-point function:

\be
W(t,x)  = <0|\Phi(t,z) \Phi(0,0)|0>\, = \sum_{m,n}
\int \frac{dp}{2\pi} \frac 1{2\omega_{p,m n}}
\re^{\ri p x - \ri \omega_{p,mn} t}\,,
\ee
where  $\omega_{p,mn}^2 = p^2 + \frac 4\theta(m+n+1) + M^2$. The expectation value
$W(t,x)$ has an analytic continuation to 
complex values in the time difference $t$ from the lower half plane $t - \ri x_4$ with $x_4 > 0$, 
\be
W(t - \ri x_4,x)  = \sum_{m,n} \int \frac{dp}{2\pi} \frac 1{2\omega_{p,m n}}
\re^{\ri p x - \ri \omega_{p,mn} ( t - \ri x_4 ) }\,.
\ee
This coincides for $t = 0$ with the Schwinger two-point function, i.e. the Euclidean correlation function:
\begin{align}
\label{correlation-euclid}
\Delta^E_{m_1n_1;m_2n_2}(x) = \delta_{m_1n_2}\delta_{m_2n_1} \int \frac{dp}{2\pi} \frac 1{2\omega_{p,m_1n_1}}
\re^{\ri p x_1 - \omega_{p,m_1n_1} |x_4|}\,,
\end{align}
for $x_4 > 0$ and with the discrete indices $m_1,\,n_1$.
This leads to the following form of the Euclidean propagator:
\begin{align}
\Delta^E_{m_1n_1;m_2n_2}(x) & = \int \frac{dP_1 dP_4}{(2\pi)^2} \frac{\delta_{m_1n_2}\delta_{m_2n_1} \re^{\ri px}}
{P_1^2+P_4^2+\frac 4\theta (m_1 + n_1+1) + M^2 }\,.
\end{align}
which incorporates the Euclidean invariance in the undeformed dimensions and has been used in previous sections.
The step back to equation \eqref{correlation-euclid} is obtained by integration over $P_4$.



\section{Conclusions}

A remarkable result of this letter concerns the appearance of the fixed point at
$\Omega=1$, at least to 1-loop, which implies the vanishing of the beta function
for the coupling constant at this parameter point:
$$
\beta_\lambda = 0.
$$
Furthermore, we observe the appearance of an additional, sixth parameter relevant in the renormalization procedure.
Its origin lies in the asymmetry between noncommutative and commutative directions at 1-loop.
We could show that this is true not only in Euclidean but also in Minkowski space-time with commuting time coordinate.
We started the program of analytic continuation of the one-loop contributions for $\Omega=1$.




\providecommand{\href}[2]{#2}\begingroup\raggedright\endgroup

\end{document}